\documentclass{article}
\usepackage{amsfonts,amssymb,amsmath,latexsym,ae,aecompl,bbm,algorithm}
\usepackage{algorithm}
\usepackage[noend]{algpseudocode}
\usepackage{graphics}
\usepackage{graphicx}
\usepackage{fullpage}

\algblockdefx[ExecBl]{BlockOn}{BlockOff}  [1]{#1}
  
\makeatletter
\ifthenelse{\equal{\ALG@noend}{t}}%
  {\algtext*{BlockOff}}
  {}%
\makeatother

\newcommand{\remove}[1]{}

\newcommand{\INDState}[1][1]{\State\hspace{4ex}}

\usepackage{amsthm}

\makeatletter
\newtheorem*{rep@theorem}{\rep@title}
\newcommand{\newreptheorem}[2]{%
\newenvironment{rep#1}[1]{%
 \def\rep@title{#2 \ref{##1}}%
 \begin{rep@theorem}}%
 {\end{rep@theorem}}}
\makeatother

\newtheorem{theorem}{Theorem}[section]
\newreptheorem{theorem}{Theorem}

\newtheorem{lemma}[theorem]{Lemma}
\newreptheorem{lemma}{Lemma}

\newtheorem{claim}[theorem]{Claim}

\usepackage{xparse}

\usepackage{blindtext}

\usepackage{morewrites}

\usepackage{wrapfig}
\usepackage{xkeyval}

\newwrite\authorbibfile%

\AtBeginDocument{%
	\immediate\openout\authorbibfile=\jobname.aub%
}%

\AtEndDocument{%
	\immediate\closeout\authorbibfile
	\InputIfFileExists{\jobname.aub}{}{}
}%

\makeatletter

\define@key{authorbib}{scale}[1]{%
	\def\AuthorbibKVMacroScale{#1}%
}

\define@key{authorbib}{wraplines}[10]{%
	\def\AuthorbibKVMacroWraplines{#1}%
}

\define@key{authorbib}{imagewidth}[4cm]{%
	\def\AuthorbibKVMacroImagewidth{#1}%
}

\define@key{authorbib}{overhang}[10pt]{%
	\def\AuthorbibKVMacroOverhang{#1}%
}

\define@key{authorbib}{imagepos}[l]{%
	\def\AuthorbibKVMacroImagepos{#1}%
}

\makeatother

\presetkeys{authorbib}{imagepos=l,imagewidth=4cm,wraplines=15,overhang=20pt}{}

\newlength{\AuthorbibTopSkip}
\newlength{\AuthorbibBottomSkip}
\NewDocumentCommand{\authorbibliography}{+o+m+m+m}{%
	\IfNoValueTF{#1}{%
	}{%
	\setkeys{authorbib}{#1}%
	\immediate\write\authorbibfile{%
		\string\begin{wrapfigure}[\AuthorbibKVMacroWraplines]{\AuthorbibKVMacroImagepos}[\AuthorbibKVMacroOverhang]{\AuthorbibKVMacroImagewidth}^^J
			\string\includegraphics[scale=\AuthorbibKVMacroScale]{#2}^^J
			\string\end{wrapfigure}^^J
	}%
}%
\IfNoValueTF{#3}{%
	\typeout{Warning: No author name}%
}{%
\immediate\write\authorbibfile{%
	\unexpanded{\vspace{\AuthorbibTopSkip}}^^J
	\string\noindent\relax
	\unexpanded{\textbf{#3}\par}^^J
	\string\noindent\relax
	\unexpanded{#4}^^J%
	\unexpanded{\vspace{\AuthorbibBottomSkip}}^^J
}%
}%
}%

\begin{document}

\title{Deterministic Backbone Creation in an SINR Network without Knowledge of Location}

\author{Dariusz~R.~Kowalski\thanks{Department of Computer Science, University of Liverpool, Liverpool, United Kingdom. D.Kowalski@liverpool.ac.uk}
	\and
	William~K.~Moses~Jr.\thanks{Department of Computer Science and Engineering, Indian Institute of Technology Madras, Chennai, India. wkmjr3@gmail.com. This work was done while the author was an intern at Xerox Research Centre India, Bangalore. }
	\and
	Shailesh~Vaya\thanks{Conduent Labs India, Bangalore, India. Shailesh.Vaya@conduent.com. This work was done while the author was with Xerox Research Centre India, Bangalore.}
	}

\date{}
\maketitle

\begin{abstract}
For a given network, a backbone is an overlay network consisting of a connected dominating set with additional accessibility properties. Once a backbone is created for a network, it can be utilized for fast communication amongst the nodes of the network.

The Signal-to-Interference-plus-Noise-Ratio (SINR) model has become the standard for modeling communication among devices in wireless networks. The SINR model takes into consideration the power of all transmitting stations as well as the relative distances between the stations to determine whether a given  transmission is successful. It is much closer to the real world and has been established to have lot more applicability.

For the SINR model, the community has pondered what the most realistic solutions for communication problems in wireless networks would look like. Such solutions would have the characteristic that they would make the least number of assumptions about the availability of information about the participating nodes. Solving problems when nothing at all is known about the network and having nodes just start participating would be ideal. However, this is quite challenging and most likely not feasible. The pragmatic approach is then to make meaningful assumptions about the available information and present efficient solutions based on this information.

We present a solution for creation of backbone in the SINR model, when nodes do not have access to their physical coordinates or the coordinates of other nodes in the network. This restriction models the deployment of nodes in various situations for sensing hurricanes, cyclones, and so on, where only information about nodes prior to their deployment may be known but not their actual locations post deployment. We assume that nodes have access to knowledge of their label, the labels of nodes within their neighborhood, the range from which labels are taken $[N]$ and the total number of participating nodes $n$. We also assume that nodes wake up spontaneously. We present an efficient deterministic protocol to create a backbone with a round complexity of $O(\Delta \lg^2 N)$.
\end{abstract}


\section{Introduction}\label{sec:intro}
 We study the backbone creation problem in Signal-to-Interference-plus-Noise-Ratio (SINR) networks, where wireless devices do not have access to information about their actual physical coordinates on the physical plane, but know the labels of their neighboring nodes. We assume availability of other standard information like the number of participating nodes $n$, the maximum degree of any node $\Delta$, and the range from which the labels of the nodes are chosen $[N]$.

  The area of SINR networks has been a source of ongoing study in recent years. It was developed as a refined alternative to the simpler radio networks model. As such the SINR model attempts to capture more aspects of wireless devices in order to more realistically model wireless networks. Radio networks typically allow a message transmitted from device $u$ to reach a neighboring device $v$ in a given round if none of $v$'s other neighbors transmit in that same round. The SINR model allows multiple devices to transmit in a given round and based upon certain mathematical formulas, dictates which devices' messages would successfully reach which other devices.

  The SINR model takes into consideration each device's transmission power and by extension the signal strength of the simultaneously transmitting stations. The ratio of a given device's signal strength at a given receiving device to the sum of the ambient noise and signal strengths of other transmitting devices at the given receiving device gives the SINR formula. Regardless of how many other devices transmit in a given round, if the related SINR formula for a transmitting device $u$ and receiving device $v$ exceeds a given threshold, then $u$'s message is received by $v$. This allows us to develop communication protocols in which multiple devices simultaneously transmit in the same round.

   In this paper, we study the problem of backbone creation for a network. It is closely related to the problem of finding a connected dominating set (CDS) amongst a set of devices, i.e. an overlay network over all the devices such that all devices are either a part of this network or are at most one hop away from some device in it. A backbone is a CDS with the following additional \textbf{properties}: (1) Each node in the CDS has a constant degree w.r.t. other nodes in the CDS, (2) the diameter of the CDS is asymptotically similar to that of the communication graph, and (3) the number of nodes in the CDS is within a constant factor of the minimum size CDS possible in the network. Computing a backbone is a fundamental problem to solve for networks. Once it is computed, it may be used as a starting point to solve other vital communication problems like broadcast, multi-broadcast, gossiping and so on.

\textbf{Our results:} We present a deterministic protocol to construct a backbone on an arbitrary connected graph with spontaneous wakeup (all nodes awake initially) in $O(\Delta \lg^2 N)$ rounds. Our algorithm works for \textbf{weak devices}, defined in \cite{JKS12,JKS-ICALP-13}, where devices must overcome additional resistance (in weak devices) in order for their messages to be received by other devices. We also adopt the notion of weak connectivity/weak links introduced in \cite{DGKN13} and refined in \cite{JK16}, in which the communication graph built on top of the network must have edges between nodes that are within range of each other instead of merely a fraction of the range, where range is dependent on the strength of the device.

\textbf{Related work:} Backbones and the closely related notion of dominating sets are well studied in the literature. For a comprehensive survey of the work done on dominating sets in various models of wireless networks, please refer to Yu et al. \cite{YWWY13}. Work has been done on dominating sets in the SINR model \cite{SRS08,YHWTL16}, where additional power is given to the devices in order to achieve good results. Scheideler et al. \cite{SRS08} use tunable collision detection to compute a dominating set which stabilizes in $O(\lg n)$ rounds with high probability. Yu et al.~\cite{YHWTL16} use power control to construct a connected dominating set in $O(\lg^2 n)$ rounds with high probability. 

  We are most interested in results pertaining to backbones, which are essentially connected dominating sets with additional constraints on the structure, in the SINR model as in \cite{JK12,CKV15,CV16,RKV15,MV16b}. The work so far has focused on weak links and weak devices. Jurdzi\'nski and Kowalski~\cite{JK12} construct a backbone in $O(\Delta \lg^3 N)$ rounds assuming spontaneous wake-up, i.e. all nodes are initially awake, and each node knows the coordinates of its location. Chlebus et al.~\cite{CKV15,CV16} construct a backbone in $O(\Delta \lg^{8a+3}N)$ rounds, where $a$ is a positive constant, with high probability assuming spontaneous wake-up. For the case of an uncoordinated start (only some nodes are awake initially), they create a backbone in $O(n \lg^2 N + \Delta \lg^{8a+3} N)$ rounds with high probability. Reddy et al.~\cite{RKV15} create a backbone when only some nodes are awake for different scenarios. They construct a backbone in $O(D \lg^2 n + k \lg \Delta)$ rounds when nodes know their own coordinates and those of their neighbors. They construct a backbone in $O(n \lg n)$ rounds when nodes only know their own coordinates. They construct a backbone in $O(n)$ rounds when nodes don't know their own coordinates but know the labels of their neighbors. Moses Jr. and Vaya~\cite{MV16b} construct a backbone in $O(n \lg^2 N)$ rounds when considering spontaneous wake-up. They construct a backbone in $O(n \lg^2 N \lg n)$ rounds from an uncoordinated start. Please refer to Table~\ref{tab:alg-comparison} for a comparison of our results and previous work.
    
  Also, in a related vein, Jurdzi\'nski et al.~\cite{JKRS15} create a quasi-backbone structure using randomization in $O(D \lg^2 n)$ rounds with high probability. A quasi-backbone is a configuration of probabilities that allows groups of nodes within certain distance of each other to communicate.

\begin{table*}[ht]
	\caption{Comparison of various algorithms that construct a backbone. $D$ is diameter of graph (based on strong/weak links), $\Delta$ is max. degree of graph, $n$ is number of nodes, $N$ is the max. value of any label of node, $a$ is a positive constant. All running times are in rounds. Note that running times of randomized algorithms are with high probability. The row with $\maltese$ is for our algorithm.}
	\begin{center}
	\resizebox{\columnwidth}{!}{%
		\begin{tabular}{|c|c|c|c|c|c|c|c|}
			\hline
			\textbf{\small Algorithm} & \textbf{\small Randomized} & \textbf{\small With} & \textbf{\small With} & \textbf{\small Device} & \textbf{\small Link} & \textbf{\small Running Time} & \textbf{\small Running Time}\\
			& & \textbf{\small Knowledge Of} & \textbf{\small Knowledge Of} & \textbf{\small Type} & \textbf{\small Type} & \textbf{\small with Spontaneous} & \textbf{\small with Uncoordinated} \\
			& & \textbf{\small Coordinates} & \textbf{\small Neighborhood} & & & \textbf{\small Wakeup} & \textbf{\small Wakeup} \\
			\hline \hline
			
			Local Learning \&  & No & Yes & No & Weak & Weak & $O(\Delta \lg^3 N)$ & - \\
			Neighborhood Learning\cite{JK12} & & & & & & &  \\
			\hline
			Local-Multicast\cite{RKV15} & No & Yes & Yes & Weak & Weak & - & $O(D \lg^2 n + k \lg \Delta)$ \\
			\hline
			Thread1, Thread2, Construct-Backbone\cite{RKV15} & No & Yes & No & Weak & Weak & - & $O(n \lg n)$\\
			\hline
			BTD$\_$Traversals, BTD$\_$MB\cite{RKV15} & No & No & Yes & Weak & Weak & - & $O(n)$\\
			\hline
			Backbone \& & Yes & No & No & Weak & Weak & $O(\Delta \lg^{8a+3}N)$ & $O(n \lg^2 N + \Delta \lg^{8a+3} N)$\\
			Emulated\_DFS\_for\_Backbone \cite{CV16}  & & & & & & &  \\
			\hline
			Backbone-Creation\cite{MV16b} & No & No & No & Weak & Weak & $O(n \lg^2 N)$ & $O(n \lg^2 N \lg n)$ \\
			\hline
			$\maltese$\textbf{Backbone-Creation} & No & No & Yes & Weak & Weak & $O(\Delta \lg^2 N)$ & -\\
			
			\hline 
		\end{tabular}
		}
	\end{center}
	\label{tab:alg-comparison}
\end{table*}

  \textbf{Organization of Paper:} This paper is organized as follows. The technical preliminaries are covered in Section~\ref{sec:prelims}. The actual backbone creation protocol is explained and analyzed in Section~\ref{sec:backbone}. The subalgorithms \emph{Leader-Election}, \emph{Two-Hop-Connection}, and \emph{Three-Hop-Connection} are described and analyzed in Sections~\ref{sec:leader-election},~\ref{sec:two-hop-connection}, and~\ref{sec:three-hop-connection} respectively. Finally, some conclusions are presented in Section~\ref{sec:conclusions}.

\section{Technical Preliminaries}\label{sec:prelims}

  A Signal-to-Interference-plus-Noise-Ratio (SINR) model of wireless networks brings the model of radio networks closer to the real world. The three fixed parameters of this model are the path loss constant $\alpha > 2$, the ambient noise $\mathcal{N} \geq 0$, and the threshold parameter $\beta \geq 1$. In a wireless network, there are several wireless stations (or devices) which are spread out in a physical plane. The Euclidean distance between two such stations $u$ and $v$ is given by distance function $d(u,v)$. Each station $u$ has its own transmission power given by $P_u$. In our setting, we assume that every device has the same transmission power. 
  
    We consider a synchronous fault free system consisting of several rounds corresponding to ticks of a global clock. In a given round, a station can either act as a receiver of messages or else a transmitter of messages but not both. If a node acted as receiver in the previous round and messages were successfully transmitted to it, then it will have access to those messages in the current round. Messages are of size $O(\lg N)$.

  In a given round, let $\mathcal{T}$ be the set of transmitting stations to which $u$ belongs. Let $v$ be a station which is in receiving mode. We define the SINR function of $u$, $v$, and $\mathcal{T}$ as follows:\\
$SINR(u,v,\mathcal{T}) = \dfrac{\frac{P_u}{d(u,v)^{\alpha}}}{\mathcal{N} + \sum \limits_{i = \mathcal{T}\backslash u} \frac{P_i}{d(i,v)^{\alpha}}}$\\

  $u$'s transmission will be successfully received by $v$ in the given round if the SINR function of the parameters crosses the threshold $\beta$, i.e.
\begin{align}\label{eq:SINR}
SINR(u,v,\mathcal{T}) \geq \beta
\end{align}

  An additional constraint that determines whether a device is \textbf{weak} or \textbf{strong} is given below.
\begin{align}\label{eq:weak-device}
\dfrac{P_u}{d(u,v)^{\alpha}} \geq (1 + \epsilon) \beta \mathcal{N}
\end{align}

  Where $\epsilon$ is a sensitivity parameter of the devices which possibly limits the effective distance that $u$'s signal can reach. Strong devices have $\epsilon = 0$ and weak devices have $\epsilon > 0$.

  For a transmitting device $u$ and a receiving device $v$ in a given time step, we say that \textbf{$u$ successfully transmits a message to $v$} (cf. \cite{JKS12,KV10}) if both conditions~\ref{eq:SINR} and \ref{eq:weak-device} are satisfied. We can also say $v$ successfully received the message from $u$. The \textbf{range} of a node $u$, denoted $r_u$, is defined as the maximum distance from $u$ that another device can be situated and still receive a message from $u$ if no other station transmits at the same time. Since every device has the same transmission power, all nodes also have the same range, denoted $r$. In a given round, if a station $u$ successfully transmits to all stations $v$ within range, then we say that \textbf{$u$ successfully transmitted in that round}.

  In order to accurately capture the communication possibilities in the network, we define a \textbf{communication graph $G(V,E)$} where the wireless devices are represented by nodes and there exists an edge from node $u$ to node $v$ if node $v$ is within range of node $u$. Henceforth, we use the terms device, station, and node interchangeably. Since every device has the same range, the graph is undirected. We assume that the graph is connected, i.e. there exists a path between any two nodes $u$ and $v$. By defining the edges of the communication graph based on nodes within range of each other, we are adopting the \textbf{weak link} model (cf. \cite{DGKN13,JK16}) of communication graphs. The \textbf{neighborhood} of a node $u$ is defined as all the nodes $v$ which have edges from $u$ in the communication graph. The \textbf{degree} of a node is defined as the size of its neighborhood. A node $v$ is said to be $k$ \textbf{hops} away from a node $u$ if there exist $k-1$ intermediate nodes in the shortest path from $u$ to $v$ on the communication graph.

  We consider nodes which \textbf{spontaneously wake up}, i.e. all nodes are initially awake and can fully participate at the start of any algorithm or protocol for the nodes.

  We overlay a 2-dimensional \textbf{grid} $G_x$ on top of the Euclidean plane on which the devices lie. Each grid box is an $x$ by $x$ square which is denoted by its bottom left coordinates. If a device has coordinates $(a,b)$, then it lies in grid box $(k,j)$ iff $kx \leq a < (k+1)x$ and $jx \leq b < (j+1)x$. The grid $G_{\frac{r}{\sqrt{2}}}$ is known as the \textbf{pivotal grid} (cf. \cite{DP07,EGKPPS09}) and has the property that all devices that lie within it are within range of each other.

  Let $k$, $m$, $N$, and $t$ be positive integers such that $1 \leq m \leq k \leq N$. A \textbf{$(k, m, N)$-selector of size $t$} (cf.~\cite{DGV03}) is a family $\mathcal{F}$ of $t$ subsets of $[N]$ such that for any subset $S$ of elements from $[N]$ of size $k$, there exist $m$ subsets in $\mathcal{F}$ that intersect with exactly one element of $S$. No two of the $m$ subsets intersect with the same element from $S$. The size of a $(k, m, N)$-selector is $O(\frac{k^2}{k-m+1} \lg N)$. When $m > k$, define a $(k, m, N)$-selector as a $(m, m, N)$-selector of size $O(m^2 \lg N)$.

  Let $N$ and $c$ be positive integers such that $N \geq c$. An \textbf{$(N,c)$-strongly selective family of size $t$} (cf. \cite{CMS01}), also referred to as an $(N,c)$-ssf of size $t$, is a family $\mathcal{F}$ of $t$ subsets of $[N]$ such that for any subset $S$ of elements from $[N]$ of size $\leq c$, for each element of $S$, there exists a subset of $\mathcal{F}$ that intersects $S$ at only that element. We choose $c$ to be a suitably large constant $\geq 441(2d+1)^2$, where $d$ comes from Lemma 2 from \cite{MV16a} and depends on the values of $\alpha$, $\beta$, $\epsilon$, and $k=21$. The size of an $(N,c)$-ssf is $O(c^2 \lg N) = O(\lg N)$.\footnotemark[1]\footnotetext{Prior known explicit constructions of an $(N,c)$-ssf take longer, while there exists existential evidence that a required construction can match our size bound. Recent research, though as yet unpublished\cite{BG15}, shows that it may be possible that an explicit construction of an $(N,c)$-ssf, where $c$ is a constant, can take $O(\lg N)$ time. This is done by constructing an $(N,(1,c-1))$ cover free family.}

  When a node executes a $(k,m,N)$ selector of size $t$ or an $(N,c)$-ssf of size $t$, the following occurs for $t$ rounds. Arrange the $t$ subsets in some total order. This order will be known to all nodes prior to the start of any protocol. The $t^{\text{th}}$ subset of the family represents the labels of nodes chosen for the $t^{\text{th}}$ round. If a node is chosen in a particular round, it performs a given action (example: transmits a message). In other rounds, the node simply acts as a receiver. We use Theorem 1 from \cite{MV16a}, restated below, extensively as a primitive in our work. In order to fully understand it, we first state Lemma 2 from \cite{MV16a}.
  
  \begin{lemma}{[Lemma~2 in \cite{MV16a}]}\label{lem:ssf-dil}
	For stations with same range $r$, sensitivity $\epsilon > 0$, and transmission power, for each $\alpha > 2$, there exists a constant $d$, which depends only on the parameters $\alpha$, $\beta$, and $\epsilon$ of the model and a constant $k$, satisfying the following property. 
	
	Let $W$ be the set of stations such that at most a constant $k$ of them want to transmit in any grid box of the grid $G_x$, $x \leq \frac{r}{\sqrt{2}}$. Let $u$ and $v$ be two stations in different grid boxes such that the distance between them, $\sqrt{2} x$, is the minimum distance between any two stations in different grid boxes in $G_x$. Let $A_u$ be the set of stations in $u$'s grid box.
	
	If $u$ is transmitting in a round $t$ and no other station within its box or a box less than $d$ box distance away from its box is transmitting in that round, then $v$ and all stations in $A_u$ can hear the message from $u$ in round $t$.
  \end{lemma}
  
  \begin{theorem}{[Theorem~1 in \cite{MV16a}]}\label{the:ssf-replace}
For a grid $G_x$, $x \leq \frac{r}{\sqrt{2}}$, let the set of all nodes that want to transmit satisfy the properties of Lemma~\ref{lem:ssf-dil}. Every node in this set can successfully transmit a message to its neighbors within $\sqrt{2}x$ distance of it in $O(\lg N)$ rounds by executing one $(N,c)$-ssf, where $c = k^2(2d+1)^2$ where $d$ is taken from Lemma~\ref{lem:ssf-dil} and $k$, a constant, is an upper limit on the number of nodes from the set in any box of the grid.
  \end{theorem}

  Each node is aware of the number of nodes in the network $n$, the range from which node labels are taken $[N]$, its own label, the labels of nodes within range of it (i.e. the node's neighborhood), and the maximum degree of any node $\Delta$. Nodes do not have knowledge of their or other nodes' coordinates in the 2-D plane. All nodes have access to information on any $(N,c)$-ssf's or $(k,m,N)$ selectors that are to be executed during the algorithm. Nodes can't perform collision detection, i.e. when a node is a receiver in a round, it is unable to distinguish between the situation where no messages are sent to it in that round and the situation where one or more messages sent to it in that round are not successfully received.

\section{Backbone Creation Algorithm and Analysis}\label{sec:backbone}
\emph{Backbone-Creation} takes a wireless network that induces a connected communication graph and produces a backbone in three stages. In the first stage, nodes execute \emph{Leader-Election} to choose at most one leader per grid box of the pivotal grid. \textbf{Helper nodes} are non-leader nodes that have been assigned to act as intermediary nodes between two leaders in the backbone. Between two leaders, there can be either one or two helper nodes, depending on whether the leaders are two or three hops away from each other respectively. In the second stage, nodes execute \emph{Two-Hop-Connection} to connect leaders two hops from each other via helper nodes. Nodes execute \emph{Three-Hop-Connection} in the third stage to connect leaders three hops from each other through helper nodes. All chosen nodes (leaders and helper nodes) together form a backbone.

\alglanguage{pseudocode}
\begin{algorithm}
\begin{algorithmic}[1]
\caption{Backbone-Creation, run by each node $u$}

\State Leader-Election
\State Two-Hop-Connection
\State Three-Hop-Connection

\Statex
\end{algorithmic}
\end{algorithm}

\begin{theorem} \label{the:backbone-creation-main}
Given a network which induces a connected communication graph, \emph{Backbone-Creation} runs in $O(\Delta \lg^2 N)$ rounds and guarantees the creation of a backbone.
\end{theorem}

\begin{proof}
We must prove that the nodes chosen to be part of the backbone form a connected dominating set (CDS) with the following additional properties

\begin{enumerate}
\item Each node in the CDS has a constant degree with respect to other nodes in the CDS.
\item The diameter of the CDS is asymptotically similar to that of the communication graph.
\item The number of nodes in the CDS is within a constant factor of the minimum size CDS possible in the network.
\end{enumerate}

First we show that a CDS is formed. Initially, at most one node (called leader) per grid box of the pivotal grid is added to the backbone through \emph{Leader-Election} and each remaining node is within the neighborhood of some leader. Thus, the set of leaders forms a dominating set. This is seen from Lemma~\ref{lem:leader-election-main}.

\begin{replemma}{lem:leader-election-main}
Given a network which induces a connected communication graph, \emph{Leader-Election} runs in $O(\Delta \lg^2 N)$ rounds and guarantees the following two conditions after completion:
\begin{enumerate}
\item Every node is either a leader or within the neighborhood of a leader.
\item No two leaders are within the neighborhoods of each other, i.e. there is at most one leader in each grid box of the pivotal grid.
\end{enumerate}
\end{replemma}

Now we need to connect the nodes in the dominating set. This is a two-step process starting with \emph{Two-Hop-Connection}, where helper nodes are added to the backbone to connect leaders two hops from each other as guaranteed in Lemma~\ref{lem:two-hop-connection-main}.

\begin{replemma}{lem:two-hop-connection-main}
After \emph{Leader-Election} is run, \emph{Two-Hop-Connection} runs in $O(\Delta \lg N)$ rounds and for each pair of leaders two hops away from each other assigns a helper node.
\end{replemma}

Now, let us consider how the graph induced by the chosen leaders and helper nodes can remain unconnected. Consider a leader $u$ with neighbor $a$ and leader $v$ with neighbor $b$. If $a$ and $b$ are neighbors but $a$ is not a neighbor of $v$ and $b$ is not a neighbor of $u$, then it is possible that $u$ and $v$ are unconnected in the CDS. This is akin to saying that $u$ and $v$ are three hops away from each other in the original communication graph. \emph{Three-Hop-Connection} is run in order to connect such leaders, as guaranteed by Lemma~\ref{lem:three-hop-connection-main}. Notice that if leaders are 4 or more hops away from each other, one of the intermediary nodes must be a leader since by Lemma~\ref{lem:leader-election-main}, all nodes must either be leaders or neighbors of leaders.

\begin{replemma}{lem:three-hop-connection-main}
After \emph{Leader-Election} is run, \emph{Three-Hop-Connect} runs in $O(\Delta \lg N)$ rounds and for every pair of leaders located three hops away from each other assigns two helper nodes.
\end{replemma}

The subgraph thus induced by the leaders and intermediary nodes forms a CDS since every node is either a leader or a neighbor of a leader and furthermore all leaders are connected to each other via intermediary nodes. Thus, after running \emph{Backbone-Creation}, we have a CDS.

Now we need to prove that the given CDS satisfies the three additional properties of a backbone. 

Property 1 states that each node in the CDS should have a constant degree with respect to other nodes in the CDS.
For a given leader, there can be at most a constant number of leaders within 2 or 3 hops of it since by Lemma~\ref{lem:leader-election-main}, there is at most one leader per grid box of the pivotal grid. By Lemma~\ref{lem:two-hop-connection-main} and Lemma~\ref{lem:three-hop-connection-main}, for every leader, the number of its neighbors in the CDS will be a constant. Every helper node can only have a constant number of leaders within range of it, since there can be at most one leader per grid box of the pivotal grid according to Lemma~\ref{lem:leader-election-main}. Therefore, even if every one of those leaders generated a constant number of helper nodes which are neighbors of the given helper node, the total number of neighbors of a helper node would still be constant. Hence Property 1 is satisfied.

Property 2 states that the diameter of the CDS should be asymptotically similar to that of the communication graph. It is clear that if the asymptotic diameter of the CDS is greater than that of the communication graph then the number of hops from one node to another increased by more than a constant factor from the communication graph to the CDS. Thus, if we show that the number of hops between any two nodes does not increase by more than a constant factor from the communication graph to the CDS, by the contrapositive, we prove Property 2. We do just this. All nodes are either leaders or one hop away from a leader by Lemma~\ref{lem:leader-election-main}. The helper nodes define the minimum hop path between leaders. Hence, the number of hops between any two nodes won't increase by more than a constant sum denoting the time taken by a node not in the backbone to transmit/receive a message to/from the backbone. Thus, Property 2 is satisfied.

Property 3 states that the number of nodes in the CDS should be within a constant factor of the minimum size CDS possible in the network. It is easy to see that the minimum size CDS possible will have more nodes than just the set of leaders generated after running \emph{Leader-Election}. We show that the number of intermediary nodes added to the backbone through \emph{Two-Hop-Connection} and \emph{Three-Hop-Connection} is only a constant times the number of leaders and thus not more than a constant factor of the minimum size CDS possible. From Lemma~\ref{lem:two-hop-connection-main} and Lemma~\ref{lem:three-hop-connection-main}, one or two helper nodes are assigned to every leader node to connect it to another leader node located two or three hops away respectively. Since there can be at most one leader per grid box of the pivotal grid by Lemma~\ref{lem:leader-election-main}, there can be at most a constant number of leaders located two hops away from a given leader and a constant number of leaders located three hops away from a given leader. Thus, totally, a constant factor of the total number of leaders is added to the network and Property 3 is satisfied.

The running time of \emph{Backbone-Creation} is the sum of the running times of \emph{Leader-Election}, \emph{Two-Hop-Connection}, and \emph{Three-Hop-Connection} and comes out to be $O(\Delta \lg^2 N)$.
\end{proof}

\section{Leader Election}\label{sec:leader-election}
\emph{Leader-Election} is run by all nodes to elect at most one leader per grid box of the pivotal grid. Furthermore, each node which isn't a leader must be within the neighborhood of a leader. 

\alglanguage{pseudocode}
\begin{algorithm}
\begin{algorithmic}[1]
\caption{Leader-Election, run by each node $u$}

\State $status \gets active$.

\For {$i \gets 0, \lceil \lg \Delta \rceil$}
	\For {Each round of execution of a $(\lceil \frac{\Delta}{2^i} \rceil + 1, \lceil \frac{41}{42}\frac{\Delta}{2^i} \rceil + 2, N)$-selector}
		\If{($u$'s degree $\in [\lceil \frac{\Delta}{21 \cdot 2^{i+1}} \rceil, \lceil \frac{\Delta}{2^i} \rceil]$) AND ($u$ is selected and none of $u$'s neighbors are) AND ($status = active$)}
			\State Execute $(N,c)$-SSF: Transmit that $u$ is a leader.
			\State $status \gets leader$
		\Else
			\State Execute $(N,c)$-SSF: Remain silent. If $u$ hears another node state that it is a leader, set $status \gets inactive$.
		\EndIf
	\EndFor
\EndFor

\Statex
\end{algorithmic}
\end{algorithm}

\begin{lemma} \label{lem:leader-election-main}
Given a network which induces a connected communication graph, \emph{Leader-Election} runs in $O(\Delta \lg^2 N)$ rounds and guarantees the following two conditions after completion:
\begin{enumerate}
\item Every node is either a leader or within the neighborhood of a leader.
\item No two leaders are within the neighborhoods of each other, i.e. there is at most one leader in each grid box of the pivotal grid.
\end{enumerate}
\end{lemma}

\begin{proof}
The running time of the algorithm can be calculated as follows. An execution of a $(\lceil \frac{\Delta}{2^i} \rceil + 1, \lceil \frac{41}{42}\frac{\Delta}{2^i} \rceil + 2, N)$-selector takes\\ $O\left( \dfrac{(\lceil \frac{\Delta}{2^i} \rceil + 1)^2}{(\lceil \frac{\Delta}{2^i} \rceil + 1 - ( \lceil \frac{41}{42}\frac{\Delta}{2^i} \rceil + 2) + 1)} \cdot \lg N \right) = O(\frac{\Delta}{2^i} \lg N)$ rounds.  From the first two for loops, we see that the total number of rounds of selectors is $\sum \limits_{i=1}^{\lceil \lg \Delta \rceil} (\frac{\Delta}{2^i} \lg N) = O(\Delta \lg N)$ rounds. Note that we'll have a constant number of executions of a $(k, m, N)$-selector at the end where $k \leq m$ and $m$ is a constant. The exact number depends on the value of $\Delta$. These selectors take $O(m^2 \lg N) = O(\lg N)$ rounds each to execute and hence don't add much to the total number of rounds of selectors executed. Now, each round of a selector is expanded into one execution of an $(N,c)$-ssf with $O(\lg N)$ rounds. Therefore \emph{Leader-Election} takes $O(\Delta \lg^2 N)$ rounds to complete.

We prove our first condition by showing that the following claim holds throughout the execution of the algorithm.
\begin{claim}\label{claim:leader-neighbor}
After the execution of the $(\lceil \frac{\Delta}{2^i} \rceil + 1, \lceil \frac{41}{42}\frac{\Delta}{2^i} \rceil + 2, N)$-selector, all nodes with degree $\in [\lceil \frac{\Delta}{2^{i+1}} \rceil, \Delta]$ will either be leaders or neighbors of leaders, $\forall 0 \leq i \leq \lceil \lg \Delta \rceil$.
\end{claim}

\begin{proof}
Initially, $i=0$ and nodes with degrees $\in [\frac{\Delta}{42}, \Delta]$ participate in the execution of a $(\Delta + 1, \lceil \frac{41}{42}\Delta \rceil + 2, N)$-selector. Our goal is find a round such that for every node with degree $\in [\lceil \frac{\Delta}{2} \rceil, \Delta]$, there exists a round of the selector such that either only the node is selected from its neighborhood or exactly one of its neighbors is selected. Then the node will either become a leader or be within the neighborhood of a leader. 

In the pivotal grid, there are at most 21 grid boxes around a given node in which its neighbor can lie. Therefore, for any node with a degree $\in [\lceil \frac{\Delta}{2} \rceil, \Delta]$, there exists at least one grid box of the pivotal grid within range of it containing at least $\lceil \frac{\Delta}{42} \rceil$ of its neighbors by pigeonhole principle. Therefore using a $(\Delta + 1, \lceil \frac{41}{42}\Delta \rceil + 2, N)$-selector ensures that there exists a round for which exactly one of the $\lceil \frac{\Delta}{42} \rceil + 1$ nodes will transmit. Thus, our claim holds for $i=0$.

Assume that the claim holds after the execution of the selectors where $i=k-1$. We want to prove that it will hold after the execution of the selector where $i=k$. At this point, all nodes with degree $\in [\lceil \frac{\Delta}{2^k} \rceil, \Delta]$ have become either leaders or neighbors of leaders and won't participate. Only nodes with degree $\in [\lceil \frac{\Delta}{21 \cdot 2^{k+1}} \rceil, \lceil \frac{\Delta}{2^k} \rceil]$ will participate. Again, in the pivotal grid, there are at most 21 grid boxes around a given node in which its neighbor can lie. Therefore, for any node with a degree in $\in [\lceil \frac{\Delta}{2^{k+1}} \rceil, \lceil \frac{\Delta}{2^k} \rceil]$, there exists at least one grid box of the pivotal grid within range of it containing at least $\lceil \frac{\Delta}{21 \cdot 2^{k+1}} \rceil$ of its neighbors by pigeonhole principle. Therefore using a $(\lceil \frac{\Delta}{2^k} \rceil + 1, \lceil \frac{41}{42}\frac{\Delta}{2^k} \rceil + 2, N)$-selector ensures that there exists a round for which exactly one of the $\lceil \frac{\Delta}{21 \cdot 2^{k+1}} \rceil$ nodes will transmit. Thus, our claim holds for $i=k$. Thus, by induction we see that our claim holds $\forall 0 \leq i \leq \lceil \lg \Delta \rceil$.
\end{proof}

After all selectors have been executed, we see that by Claim~\ref{claim:leader-neighbor}, all nodes with degree $\in [1, \Delta]$ have become either leaders or neighbors of leaders.

The proof of the second condition is simple. Let two nodes be neighbors and possible candidates for leaders. Let both of them have rounds in some selectors (need not be the same selector) where they are the sole transmitters from their neighborhoods. It is not possible for both of them to become leaders because there is a total order on the sequence of such rounds of selectors. As soon as one node gets its chance to transmit, the other node will become inactive and stop being a candidate for being a leader.
\end{proof}

\section{Two Hop Connection}\label{sec:two-hop-connection}
\emph{Two-Hop-Connection} is run by all nodes to choose helpers for leaders located two hops away from each other. \emph{Two-Hop-Connection} first calls \emph{Neighborhood-Inform} to spread information about leaders' neighborhoods to surrounding non-leaders and then has these non-leaders decide who becomes a helper node.

\alglanguage{pseudocode}
\begin{algorithm}
\begin{algorithmic}[1]
\caption{Neighborhood-Inform, run by each node $u$}

\For {$i \gets 1, \Delta$}
	\If {$u$ is a leader}
		\State Execute $(N,c)$-SSF: Transmit the label of $u$'s $i^{\text{th}}$ neighbor if $u$ has one, else do nothing.
	\Else
		\State Execute $(N,c)$-SSF: Remain silent. For every leader $v$ heard, maintain list of nodes in their neighborhood $N_v$.
	\EndIf
\EndFor

\Statex
\end{algorithmic}
\end{algorithm}

\begin{lemma} \label{lem:neighborhood-inform-main}
After \emph{Leader-Election} is run, \emph{Neighborhood-Inform} runs in $O(\Delta \lg N)$ rounds and guarantees that for any node $u$ in the neighborhood of a leader $v$, $u$ will learn the list of all neighbors of $v$, $N_v$.
\end{lemma}

\begin{proof}
The running time is calculated as the time to run $O(\Delta)$ $(N,c)$-SSFs, each of $O(\lg N)$ rounds. In any given round of the $(N,c)$-ssf, at most one leader will transmit from every box of the pivotal grid. Thus, by Theorem~\ref{the:ssf-replace} , for a given leader, all its neighbors will receive its messages.
\end{proof}

\alglanguage{pseudocode}
\begin{algorithm}
\begin{algorithmic}[1]
\caption{Two-Hop-Connection, run by each node $u$}

\State Neighborhood-Inform
\For {Each round of an $(N \cdot N,c \cdot c)$-SSF defined on pairs of labels $(s,t) \in N \times N$}
	\If {(Pair $(s,t)$ is selected such that $s$ and $t$ are both leaders in $u$'s neighborhood) AND ($u$ is the minimum label in $N_s \bigcap N_t$)}
		\State Transmit that $u$ is a helper node to leaders $s$ and $t$.
	\Else
		\State Remain silent. Store info on any nodes claiming to be $u$'s helper nodes to other leaders.
	\EndIf
\EndFor

\Statex
\end{algorithmic}
\end{algorithm}

\begin{lemma} \label{lem:two-hop-connection-main}
After \emph{Leader-Election} is run, \emph{Two-Hop-Connection} runs in $O(\Delta \lg N)$ rounds and for each pair of leaders two hops away from each other assigns a helper node.
\end{lemma}

\begin{proof}
The running time of an $(N \cdot N,c \cdot c)$-SSF is $O(c \cdot c \lg (N \cdot N)) = O(\lg N)$ rounds. The running time of \emph{Two-Hop-Connection} is therefore decided by \emph{Neighborhood-Inform}, for a running time of $O(\Delta \lg N)$ rounds.
  By Lemma~\ref{lem:neighborhood-inform-main}, each non-leader knows the neighborhoods of its surrounding leaders. For a given non-leader, there are at most a constant number of leaders in its neighborhood and by extension there are a constant number of pairwise selections of leaders in its neighborhood. Therefore by using an $(N \cdot N,c \cdot c)$-SSF and Theorem~\ref{the:ssf-replace}, leaders two hops away from each other will hear about a given helper node. Furthermore, for any two leaders with a shared neighborhood, there is exactly one non-leader node with the least label in that shared neighborhood. Thus, there is exactly one helper node that will be assigned to every pair of leaders that are two hops from each other.
\end{proof}

\section{Three Hop Connection}\label{sec:three-hop-connection}
\emph{Three-Hop-Connection} is run by all nodes to choose helper nodes for leaders three hops away from each other. As a part of \emph{Three-Hop-Connection}, nodes participate in a round-robin token passing scheme \emph{Token-Passing} where a leader coordinates the passing of a ``token" to all its neighbors. Here, when a node $u$ ``passes a token" to a node $v$, $u$ is transmitting a message saying $v$ can be active. Note that the order on neighbors arises from the total order on labels of nodes.

\alglanguage{pseudocode}
\begin{algorithm}
\begin{algorithmic}[1]
\caption{Token-Passing(Message $msg$), run by each node $u$}

\If {$u$ is a leader}
	\State Generate a token.
	\For {$i \gets 1, \Delta$}
		\State Execute $(N,c)$-SSF: Stay silent.
		\State Execute $(N,c)$-SSF: Pass token to $u$'s $i^{\text{th}}$ neighbor if it exists, else keep it.
		\State Execute 2 $(N,c)$-SSFs: Stay silent.
	\EndFor

\Else
	\For {$i \gets 1, 2 \Delta$}
		\If {$u$ receives token from leader(s)}
			\State Execute $(N,c)$-SSF: Transmit $msg$.
			\State Execute $(N,c)$-SSF: Pass token(s) back to leader(s) who sent them. 
		\Else
			\State Execute $(N,c)$-SSF: Store any messages $u$ hears.
			\State Execute $(N,c)$-SSF: Stay silent and listen for token.
		\EndIf
	\EndFor
\EndIf

\Statex
\end{algorithmic}
\end{algorithm}

\begin{lemma} \label{lem:token-passing-main}
After \emph{Leader-Election} is run, \emph{Token-Passing} runs in $O(\Delta \lg N)$ rounds and guarantees that any message transmitted by a node will be successfully received by its neighbors.
\end{lemma}

\begin{proof}
The running time can be seen as $O(\Delta)$ executions of an $(N,c$)-SSF which takes $O(\lg N)$ rounds. We now need to show that any message transmitted by a node will successfully be received by its neighbors. Note that only nodes with tokens transmit a message. We will show that at most 21 tokens can be present in the same grid box of the pivotal grid during the execution of a given $(N,c)$-SSF. Then by Theorem~\ref{the:ssf-replace}, and our choice of $c=441(2d^2 + 1)$, every node which participates in the $(N,c)$-SSF and transmits a message will have that message successfully received by its neighbors.

Consider an arbitrary node. There are at most 21 grid boxes of the pivotal grid which are within range of it. There can be at most one leader per grid box by Lemma~\ref{lem:leader-election-main} and there is only one token associated with each leader. Suppose all these leaders pass their tokens to the grid box in which the node is present. Then there can be at most 21 nodes participating in a given $(N,c)$-SSF from the same grid box of the pivotal grid. 
\end{proof}


\alglanguage{pseudocode}
\begin{algorithm}
\begin{algorithmic}[1]
\caption{Three-Hop-Connection, run by each node $u$}

\State Message $msg$. Here $msg$ is the message that the node wants to transmit.

\If {$u$ is not a leader}
	\State $msg \gets$ $u$'s label and labels of any leaders in $u$'s neighborhood
\EndIf

\State Token-Passing($msg$)

\If {$u$ is not a leader}
	\For{Every leader, $b$, $u$ hears about}
		\State If $u$ has heard from multiple nodes $y_1, y_2, \ldots, y_{\Delta}$ about a given leader $b$, select the $y_i$ with minimum label that belongs to $b$.
	\EndFor
	\State $msg \gets$ $u$'s label and any pairs $(y_i, b)$ that $u$ has chosen
\EndIf

\State Token-Passing($msg$)

\If {$u$ is a leader}
	\For{Every other leader, $b$, $u$ hears about}
		\If {($u$ has heard from nodes $x_1, x_2, \ldots$ about leader $b$) AND ($u$ doesn't have a helper node to $b$)}
			\State The intermediate nodes between $u$ and $b$ will form pairs $<x_i, y_i>$. 
						 Choose the smallest label from all $x_i$'s and $y_i$'s. 
			\If {Smallest label is $x_s$}
				\State Choose corresponding lowest label $y_s$ amongst all pairs $<x_s, y_i>$. 
			\ElsIf {Smallest label is $y_s$}
				\State Choose corresponding lowest label $x_s$ amongst all pairs $<x_i, y_s>$. 
			\EndIf
		\EndIf
	\EndFor
	\State $msg \gets$ $u$'s label and triplets $(x_s, y_s, b)$ that $u$ has chosen
\EndIf

\State Execute $(N,c)$-SSF: If leader then transmit $msg$, else store info $u$ hears about itself.

\Statex
\end{algorithmic}
\end{algorithm}

\begin{lemma} \label{lem:three-hop-connection-main}
After \emph{Leader-Election} is run, \emph{Three-Hop-Connect} runs in $O(\Delta \lg N)$ rounds and for every pair of leaders located three hops away from each other assigns two helper nodes.
\end{lemma}

\begin{proof}
The running time of \emph{Three-Hop-Connect} is essentially the running time of \emph{Token-Passing} and by Lemma~\ref{lem:token-passing-main}, it's $O(\Delta \lg N)$. We need to prove that for every pair of leaders located three hops away from each other, they both choose the same set of helper nodes. We also need to show that once they transmit this information, their neighbors successfully hear the message. We also need to ensure that the messages being passed to \emph{Token-Passing} are of size $O(\lg N)$ bits and not more (in order not to break our assumption that messages are of size $O(\lg N)$).

First we show that the size of messages being passed to \emph{Token-Passing} are of $O(\lg N)$ bits. The first message contains information about each of the leaders in the neighborhood of a node. There are at most a constant number of leaders in the neighborhood of a given node because at most one leader exists per grid box of the pivotal grid by Lemma~\ref{lem:leader-election-main}. Hence the first message is of $O(\lg N)$ bits. The second message contains information about leaders located at most 3 hops away from a given node. Again, by the same logic, this is a constant number and hence the second message is of size $O(\lg N)$. Therefore all messages being passed to \emph{Token-Passing} are of size $O(\lg N)$.

Now we show that for any pair of leaders $u$ and $v$ three hops away from each other, they eventually choose the same helper nodes $a$ and $b$. Let $a$ be the helper node with the lowest label among all possible 2 hop paths from $u$ to $v$ and from $v$ to $u$. $b$ is the helper node with the lowest label among possible helper nodes connecting $a$ to the other leader. $a$ and $b$ are clearly unique. Without loss of generality, let $a$ be in the neighborhood of $u$ and let $b$ be in the neighborhood of $v$. Therefore, in the second message sent through \emph{Token-Passing}, information about $b$ and $v$ will be sent to $u$ through $a$ and similarly information about $a$ and $u$ will be sent to $v$ through $b$. Thus, both leaders will decide on $a$ and $b$. 

Now we show that all leaders will successfully inform their neighbors about helper nodes. Each leader has chosen its helper nodes to the constant number of other leader nodes three hops away from it. Thus, the message that will be transmitted will be of size $O(\lg N)$ bits. Now, only leaders participate in the final $(N,c)$-SSF execution and there is at most one leader per grid box of the pivotal grid. Therefore by Theorem~\ref{the:ssf-replace}, all leaders are able to successfully transmit their messages to their neighbors.
\end{proof}

\section{Conclusions}\label{sec:conclusions}
In this paper, we have constructed a backbone for an SINR network when knowledge of each node's coordinates is missing and instead knowledge of the immediate neighbors of each node is supplied. Backbone creation has not been dealt with before for this particular model and we provide such a protocol which solves this problem in $O(\Delta \cdot \lg^2 N)$ rounds. This work shows how to extend the technique used in \cite{MV16a} to the setting where knowledge of neighbors is known and uses this extra information to derive better bounds on the time taken to create a backbone.


\bibliographystyle{abbrv}
\bibliography{reference}

\end{document}